\input epsf
\epsfverbosetrue
\documentstyle{l-aa}

\begin{document}
\newcommand{\ana}{A\&A~}  
\newcommand{\ans}{A\&AS~}  
\newcommand{\apj}{{ApJ~}}     
\newcommand{\mnras}{{MNRAS~}} 
\newcommand{\ea}{{et~al.~}}
\newcommand{\eg}{{e.g.~}}
\newcommand{\angs}{\AA~}
\newcommand{\cggs}{$10^{-9}$ ergs cm$^{-2}$ s$^{-1}$ sr$^{-1}$ \AA$^{-1}$~}
\newcommand{\cgs}{ergs cm$^{-2}$ s$^{-1}$ sr$^{-1}$ \AA$^{-1}$}

\thesaurus{02 (11.05.2; 11.06.2; 12.03.3; 12.04.2) }
\title{Extragalactic backround light: the contribution by faint and
low surface brightness galaxies}
\author{Petri V\"ais\"anen\thanks{{\it Present address:\/} Harvard-Smithsonian
Center for Astrophysics, 60 Garden St., Cambridge, 02138 MA, USA -- e-mail:
pvaisanen@cfa.harvard.edu}}
\institute{Observatory, P.O. Box 14, FIN-00014 University of Helsinki, Finland}
\date{received: orig. Feb. 1996; accepted: April 16, 1996}
\maketitle

\begin{abstract}
Several models which have been
constructed to explain the faint galaxy excess in observed number counts
are used to predict the intensity of the extragalactic background light (EBL).
Special attention is given to
irregular and dwarf galaxies, which seem to be more common in the universe
than once
thought, and to low surface brightness galaxies (LSB), which
can in principle be altogether missed from galaxy counts.
The nature of the latter objects is still
unclear, but some plausible models predict
that LSB galaxies can
increase the intensity of the EBL by a factor of up to 5 from a standard,
no-evolution model in the optical and near infrared
and by an order of magnitude in the UV.  If the faint
excess population consists of low-luminosity dwarfs, whose luminosity
function has a steep faint end, the EBL can well increase by a factor of
3 to 5, while still being consistent with current number count data.
The resulting values
of the EBL are not far from the observed upper limits.
In the future the overall level of the EBL and its spectral distribution could
be used to differentiate between galaxy population models.

\keywords{Diffuse radiation -- Cosmology: observations --  Galaxies:
evolution -- Galaxies:
luminosity function}
\end{abstract}

\section{Introduction}

The importance of the extragalactic background light (EBL)
for cosmology has long been recognized.  This integrated
diffuse background radiation in the
optical, ultraviolet and infrared wavebands
contains information about otherwise difficult-to-observe or completely
unobservable periods of the universe's past, particularly the era of
galaxy formation.  The EBL may also be useful in discriminating between
cosmological models.  For a
review of the history of the subject, see Harrison (1990);
and for both cosmology and galaxy evolutionary effects see
Partridge \& Peebles (1967) and the many papers by Tinsley
(\eg 1973, 1977).

In observational cosmology
the nature of a background brightness measurement has in principle
an advantage over the
number count observations.  When counting galaxies, whether in magnitude
or redshift bins, one needs to consider many kinds of selection effects which
affect the completeness of the sample.  Measurements of the EBL are not
plagued by this particular problem.  However, so far the
EBL has not had much success as a cosmological probe or as a tool to
investigate the evolution and origin of galaxies.  This is because the
accurate elimination of the foreground components of the sky brightness
has proved to be difficult and we
lack a generally accepted measured value of the
EBL (for a review of observational status see Mattila \ea 1991).  And even if
we had such a measurement, it would still not be easy to disentangle
the roles of cosmological parameters, galaxy evolution and
luminosity functions of galaxies (\eg Tinsley 1973).

Although this work concentrates on optical wavelengths the
treatment is essentially the same in the IR and UV
(Franceschini \ea 1991; Lonsdale 1995; Jakobsen 1995) which have
recently been more active research areas than than the optical EBL.
In the IR (see Franceschini \ea 1991; Hauser 1995) analysis is currently being
carried out on data from the Diffuse Infrared Background Experiment (DIRBE) on
board COBE.  In the near future there will be additional EBL measurements  in
the IR by Infrared Space Observatory (ISO) and later by the Space Infrared
Telescope Facility (SIRTF).
In the UV, recent observations and arguments
by Sasseen \ea (1995) indicate that the component formerly interpreted as
extragalactic is in fact produced by galactic cirrus.  Their method
utilized the power spectrum of background light; the result
implies that the  {\em optical} background fluctuations detected
by Shectman (1973, 1974), using the same method, were also galactic in origin.

During the past years, much of observational cosmology has focused
on deep galaxy-counts reaching ever fainter limits.  This in turn
has produced new models for the galaxy population.
The EBL has not generally been used as a further
constraint on these models, because of the
difficulties mentioned above.  However, it
continues to be  a vital part of observational cosmology,
especially in anticipation of near-future IR space observations.

The well-known apparent excess in the
number counts of galaxies at faint magnitudes, most notably in the $B$-band,
has led to numerous
investigations as to the nature of this effect (\eg Ferguson \& McGaugh
1995, FMG95, and references therein).
Over the years the discrepancy has been
between the counts and the predictions made
by using standard
cosmology and no, or very modest, galaxy evolution.  The suggestions
for solving the puzzle have
included altering either the cosmology (\eg introducing a
non-zero cosmological constant)
or the galaxy population (\eg introducing new galaxy populations or
altering the properties of giant galaxies via
density or luminosity evolution.)

In the past year observations of faint galaxies have led to advances in the
understanding to the faint excess and galaxy evolution.
The HST Medium Deep Survey (MDS) provided evidence which indicates
that the blue excess in number counts results from an excess population
(relative to standard Hubble class -mixes) of late-type/irregular galaxies
(Glazebrook \ea 1995; Driver \ea 1995a, 1995b; Casertano \ea 1995).  Early
results from the Hubble Deep Field also support the same conclusion
(Abraham \ea 1996).
The Canada-France redshift survey (Lilly \ea 1995a, 1995b) has also provided
new data: results show a nearly unevolving
early type population and a brightening LF of bluer galaxies.
The mild evolution of elliptical galaxies was also found in an analysis of
HST data (Im \ea 1996).
Finally, Cowie \ea (1995) announced evidence of
massive galaxies forming in the redshift range $z=1-2$ and Steidel \ea (1996)
at redshifts $z>3$.

In recent years there has also been cumulative
evidence for a significant
population of galaxies with very low surface brightnesses (\eg
Schombert \ea 1992; de Blok \ea 1995; Davies \ea 1988; for a detailed
rewiew of
the field see especially
McGaugh 1995 -- hereafter MG95);
i.e.\ surface brightnesses comparable to or fainter than the level of the
night sky.
It has been argued that the LSB population could actually {\em
be}, at least partially,
the local counterpart of the faint blue population (McGaugh 1994).
In this line of thought, a population of intrinsically LSB galaxies
would have gone undetected in the local
galaxy surveys due to selection effects;
at the same time they would be more easily detected at larger
distances in deep counts (which have much lower isophotal limits).
There has been some work recently on quantifying the
effect of observational selection criteria on the properties of observed galaxy
populations; see Davies 1990; Yoshii 1993; Davies \ea 1994; McGaugh \ea 1995;
FMG95; MG95.

The presence of LSB galaxies
affects many areas of extragalactic astronomy.
In particular, the luminosity functions hitherto derived from the local
observable galaxies have a strong underrepresentation of LSB galaxies.
As MG95 points out, even a small number of observed
LSB galaxies implies a large underlying population because of the small volume
sampling when detecting them.  There may also
be a large population
of dwarf galaxies escaping the magnitude limits of present surveys.
The goal of this work is to quantify the effect of faint and
low-surface-brightness galaxies on the extragalactic background light
and to examine whether or not existing observational limits of the
EBL constrain any
proposed models of faint galaxy properties.  In addition, the
basic ingredients which affect the surface brightness of the
extragalactic component of the sky are reviewed.

\section{Model construction}
\label{mod}

A Friedmann-Lema$\hat i$tre-Robertson-Walker universe is assumed with
the possibility for a non-zero cosmological constant $\lambda$.
The  equations describing the cosmological geometry for a $\lambda \neq 0$
universe are derived and presented in numerous
articles and textbooks, e.g.\
Weinberg (1972).  For completeness a summary of the important equations for a
general cosmology is included in Appendix~\ref{app}.

Three different cosmologies are adopted,
$(\Omega_{0},\lambda_{0}) = (0.1,0.9), \; (0.1,0)$ and $(1.0,0)$,
referred to as cases A, B and C, respectively (Table~\ref{cosmo}).
Unless otherwise stated, we use $H_{0} = 50$ Mpc km$^{-1}$ s$^{-1}$.
Note that for the {\em no-evolution models}
the value of $H_{0}$ does not affect the number counts or the EBL --
$H_{0}^{3}$ in the luminosity function cancels out with the $H_{0}^{-3}$
dependance of the
volume-element (see Eqs.~(5), (4), (A2), and (A4) below)
and the effect on the distance modulus (Eqs.~(1) and (A3))
is cancelled by the opposite dependance of $M^{\ast}$.

\begin{table}
\begin{center}
\caption{Cosmologies used, with $H_{0}=50$ Mpc km$^{-1}$ s$^{-1}$}
\begin{tabular}{lrr}
\hline
\hline
 & $\Omega_{0}$ & $\lambda_{0}$ \\
\hline
A & 0.1 & 0.9 \\
B & 0.1 & 0.0 \\
C & 1.0 & 0.0 \\
\hline
\end{tabular}
\end{center}
\label{cosmo}
\end{table}

\begin{figure}
     \epsfbox{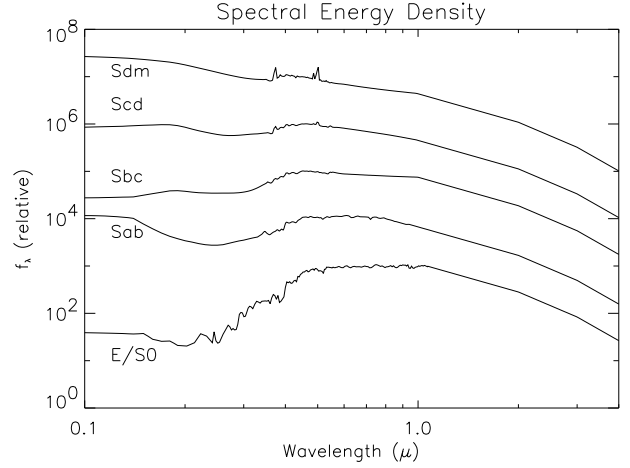}
     \caption{Spectral energy densities of five galaxy types.}
     \label{seds}
\end{figure}

The apparent magnitude $m_{\lambda}$ in a waveband centered at $\lambda$
of a galaxy of absolute magnitude
$M_{B_{J}}$ (number count data is often in Tyson's $B_{J}$ band)
at redshift $z$ is given by
\begin{equation}
\label{magnit}
m_{\lambda} = M_{B_{J}}(0,t_{0}) +
5 \log \left( \frac{d_{L}}{10 \ {\rm pc}}\right) + K_{\lambda}(z) +
E_{\lambda}(z) + C_{\lambda}
\end{equation}
where $d_{L}$ is the luminosity distance and $K_{\lambda}(z)$, the
K-correction, accounts for the redshift of the galaxy spectrum and the
stretching of the band pass:
\begin{eqnarray}
K_{\lambda}(z) & = & M_{\lambda}(z,t_{0}) - M_{\lambda}(0,t_{0})
\nonumber \\
& = & -2.5 \log \frac{\int_{0}^{\infty} f(\lambda^{'}/(1+z),0) \,
R_{\lambda}(\lambda^{'}) \, d\lambda^{'}} {\int_{0}^{\infty} f(B_{J},0) \,
R_{B_{J}} (\lambda^{'}) \, d\lambda^{'}}  \\
&   & + 2.5\log(1+z) \nonumber,
\end{eqnarray}
where $f(\lambda,z)$ is the observed spectral energy density of a
given galaxy at redshift $z$ at wavelength
$\lambda$ and $R_{\lambda}(\lambda^{'})$ are the transmission functions
of the filter bands for the $B_{J}$ band and the
Johnson UBVRIJK system (Fukugita \ea 1995).
The SEDs are identical to those of Yoshii \&
Takahara (1988; YT88) except that for the UV-spectrum of ellipticals the
``UV-intermediate'' case (see Yoshii \& Peterson 1991) is adopted,
represented by the SED of NGC 3379.
There are five different types of SED's: E/S0, Sab, Sbc, Scd, and
Sdm (see Figure~\ref{seds}) with a corresponding mix of
35:20:25:10:10 percent (\eg Tinsley 1980; Peterson \ea 1986; Guiderdoni
\& Rocca-Volmerange 1990).

$E_{\lambda}(z)$ is the correction factor for evolution:
\begin{eqnarray}
E_{\lambda}(z) & = & M_{\lambda}(z,t(z)) - M_{\lambda}(z,t_{0}) \nonumber \\
& = & -2.5 \log \frac{\int_{0}^{\infty} f(\lambda^{'}/(1+z),z) \,
R_{\lambda}(\lambda^{'}) \, d\lambda^{'}} {\int_{0}^{\infty}
f(\lambda^{'}/(1+z),0) \, R_{\lambda}(\lambda^{'}) \, d\lambda^{'}}.
\end{eqnarray}
This brings in the only dependance on $H_{0}$ through the calculation of $t(z)$
(relation (A2)).
Finally the $C_{\lambda} \equiv M_{\lambda}-M_{B{J}}$ terms are constants
which define the zero
point of the magnitude system; these are found in Table~\ref{scales} and are
adopted from Henden and Kaitchuck (1982) Section 2.5.

\begin{table}
\begin{center}
\caption{Photometric constants for determining the zero point of the
magnitude systems.  For $B_{J}$ the flux at 0.0 mag is $5.65 \cdot
10^{-9}$ ergs cm$^{-2}$ s$^{-1}$ \AA$^{-1}$~ .}
\begin{tabular}{lr}
\hline
\hline
Band & $C_{\lambda}$ \\
\hline
$U$ & 0.28 \\
$B$ & -0.17 \\
$B_{J}$ & 0.00 \\
$V$ & 0.39 \\
$R$ & 1.27 \\
$I$ & 2.08 \\
$J$ & 3.08 \\
$K$ & 5.38 \\
\hline
\end{tabular}
\end{center}
\label{scales}
\end{table}

\epsfxsize=9cm

The luminosity function is assumed to have the usual Schechter form
(Schechter 1976), expressed in magnitudes:
\begin{eqnarray}
\label{schecter2}
\Psi(M) dM & = & 0.92 \phi^{\ast} \exp \lbrace -0.92(\alpha +1)(M-M^{\ast})
- \\
 & & \exp \lbrace -0.92(M-M^{\ast}) \rbrace \rbrace dM. \nonumber
\end{eqnarray}

Values from Efstathiou \ea (1988) are adopted as the standard set of
LF parameters:
$M_{{B}_{J}}^{\ast} = -19.6 + 5 \log h$, $\alpha = -1.1$, and
$\phi^{\ast} = 1.6\cdot
10^{-2} \, h^{3}$ (though it varies slightly in the following models to get
a consistent overall normalization).
These are maximum-likelihood va\-lues derived from a set
of different surveys.  Given the present uncertainty in the normalization  of
the local population of galaxies, these values are not significantly different
from more recent LF determinations.
The Schechter parameters for additional populations are considered with
each case separately and are summarized
in Table~\ref{lumfun}.

\begin{table*}
\caption{The luminosity function parameters for various models.
The values correspond to $H_{0}=50$ Mpc km$^{-1}$ s$^{-1}$.
a) The normal giant population,  with (PLE) or without (NE) standard,
pure luminosity
evolution; b) The {\em extra}
populations of galaxies added to a {\em non-evolving} giant population
For giants the LF
has the same shape in all models with normalization:
$\varphi^{\ast}=2.0\cdot 10^{-3}$ in DRs and
$\varphi^{\ast}=2.3\cdot 10^{-3}$ in EDP.  DR1 is a non-evolving
dwarf-rich model from Driver \ea (1994) and DR2 is its
modified version; EDP has an evolving dwarf population included, with
characteristics of a single star-burst.  c)  The {\em extra}
populations of galaxies added to a passively {\em evolving}
giant population. Giant normalization:
$\varphi^{\ast}=1.7\cdot 10^{-3}$ in LZ2 and
$\varphi^{\ast}=2.0\cdot 10^{-3}$ in BBG.  In LZ2 the
giant $z_{\rm for}=2$, in all others $z_{\rm for}=5$.
d) Modified characteristics of giant population.  In BBG the later type
giants have an extra brightening over the passive luminosity evolution.  The
LF of FMB is taken from Ferguson \&
McGaugh (1995)}
\begin{center}
\begin{tabular}{l|lll|l}
\hline
\hline\noalign{\smallskip}
Model & $M_{B_{J}}^{\ast}$ & $\alpha$  & $\varphi^{\ast}$  & Notes \\
\noalign{\smallskip}
\hline
\noalign{\smallskip}
a)  & \multicolumn{3}{|l|}{ } \\
NE/  & & & &  passive lum. \\
PLE  & $-21.1$  &  $-1.1$ &  $2.4/1.7 \cdot 10^{-3}$ Mpc$^{-3}$ &
evolution  \\
\noalign{\smallskip}
\hline\noalign{\smallskip}
b) & \multicolumn{3}{|l|}{ } & \\
DR1   & $-18.0$  & $-1.8$ & $4.0\cdot 10^{-3}$ & dE (SED of E/S0)\\
      & $-18.0$ & $-1.8$ &  $8.0\cdot 10^{-3}$  & dI (SED of Sdm) \\
DR2   & $-18.0$ & $-1.5$ & $1.2\cdot 10^{-2}$   & flat SED \\
EDP   & $-19.1+2.5\lg(1/(6z+1))\,$ ;$z \leq 1.2$   & $-(1.05+z)^{2}$ ;$z \leq
0.25$ & $2.0\cdot 10^{-3}$ & flat SED \\
      & $-22.8+2.5\lg(1/(6z+1))\,$ ;$z > 1.2$ &  $-1.7\,$ ;$z > 0.25$ &
$2.0\cdot 10^{-3}$    &  \\
\noalign{\smallskip}
\hline\noalign{\smallskip}
c) & \multicolumn{3}{|l|}{ } & \\
LZ2  & $-19.1$ & $-1.8$ & $7.5\cdot 10^{-3}\cdot (1+0.5/z)^{-1}$  & SED of Sdm
\\
BBG  & $-19.1$ & $-1.8$ & $4.0\cdot 10^{-3}$  & SED of Sdm\\
\noalign{\smallskip}
\hline\noalign{\smallskip}
d)  & \multicolumn{3}{|l|}{ } & \\
BBP   & $-20.6-1.2 z\,$ ; $z \leq 1.0$ &    &   & for Scd and \\
      & $-21.8\,$ ; $z \leq 1.0$            &    &   & Sdm  \\
FMB & \multicolumn{3}{|l|}{The
LF is taken from Ferguson and
McGaugh 1995, Table 1., model B} & \\
\noalign{\smallskip}
\hline
\end{tabular}
\end{center}
\label{lumfun}
\end{table*}

Using the luminosity function and the co-moving volume element, and
integrating them over redshift, we get the number of galaxies $N(m) \ dm$
per steradian with apparent magnitude $m$ to $m+dm$:
\begin{equation}
\label{number}
N(m_{\lambda}) \, dm = \int_{0}^{z_{\rm for}} \sum_{i}
\Psi_{i}(M_{\lambda},z)
\frac{dV}{d\omega dz} \, dm \, dz,
\label{nrocount}
\end{equation}
where $\Psi_{i} (M_{\lambda},z)$ is the galaxy type -dependent (and
possibly redshift-dependent) luminosity function where
$M_{\lambda} = M_{B_{J}}$ is given by Equation~\ref{magnit}.

Integrating the number counts and the flux $f(m)$,
\begin{equation}
\label{flux}
f(m_{\lambda}) = 10^{-0.4(m+20.62-C_{\lambda})} \, {\rm ergs} \, {\rm cm}^{-2}
{\rm s}^{-1}
{\rm str}^{-1} {\rm \AA}^{-1},
\end{equation}
over the apparent magnitudes, we get the
background light per steradian contributed by galaxies:
\begin{equation}
\label{backsurf}
I_{\rm EBL} = \int_{m_{\rm cut}}^{\infty} N(m) f(m) \, dm \ {\rm ergs}\,{\rm
cm}^{-2} {\rm s}^{-1}
{\rm str}^{-1} {\rm \AA}^{-1}.
\label{eblflux}
\end{equation}
A unit of \cggs is used for $I_{\rm EBL}$ throughout this paper.

First, a cosmological world model and a galaxy evolution model is selected,
then the number counts are calculated using Eq.~(\ref{nrocount}) and
the background intensity using Eq.~(\ref{eblflux}).
A value of
$z_{\rm for} = 5$ is used
for the redshift of galaxy formation unless otherwise stated.
The limiting magnitude above which galaxies are unresolved, and
thus contribute to the background, is denoted by $m_{\rm cut}$.

The Lyman discontinuity at $\lambda_{Ly}=912$ \angs enters the filter
centered at
$\lambda$ after $z_{\rm lim} = \lambda /\lambda_{\rm Ly} - 1$.
The galaxies are assumed to be
completely opaque to Lyman continuum photons (gas absorption),
i.e.\ the integral in
Eq.~(\ref{nrocount}) is cut off at $z=z_{\rm lim}$, so in this
sense the derived $I_{\rm EBL}$ is a lower limit of EBL.
This cut-off affects mainly
the $U$-band
($z_{\rm lim} = 2.9$) and only very little the $B_{J}$ band
($z_{\rm lim} = 4.0$, beyond which very little EBL would
be coming in any case).  The effect is larger for models with luminosity
evolution (bright early stages of galaxies)
and open or $\lambda$-dominated cosmologies.
Effects due to dust absorption in galaxies and gas or dust absorption
in intergalactic space are not treated at all in this work.
Opacities, whether inside galaxies or outside, are not a
straighforward issue, either observationally or theoretically; see
Leroy \& Portilla (1996) and references therein
for a discussion of the effects of galactic opacities on number counts
and Yoshii \& Peterson (1994) and Madau (1995) for effects arising from
intergalactic absorption.

An additional cut-off magnitude is
needed at the faint end of the luminosity function.
A practical limit of current surveys does not allow the LF
to extend
much beyond a level of 4 mag fainter than $M_{B_{J}}^{\ast}$, but we
extrapolate the faint end to 8 mag fainter than $M_{B_{J}}^{\ast}$
(Yoshii 1993).
This assumption most strongly affects
those models with steep LF slopes ($\alpha \leq -1.5$).

\section{Faint galaxy populations and EBL}

\begin{figure*}
     \epsfbox{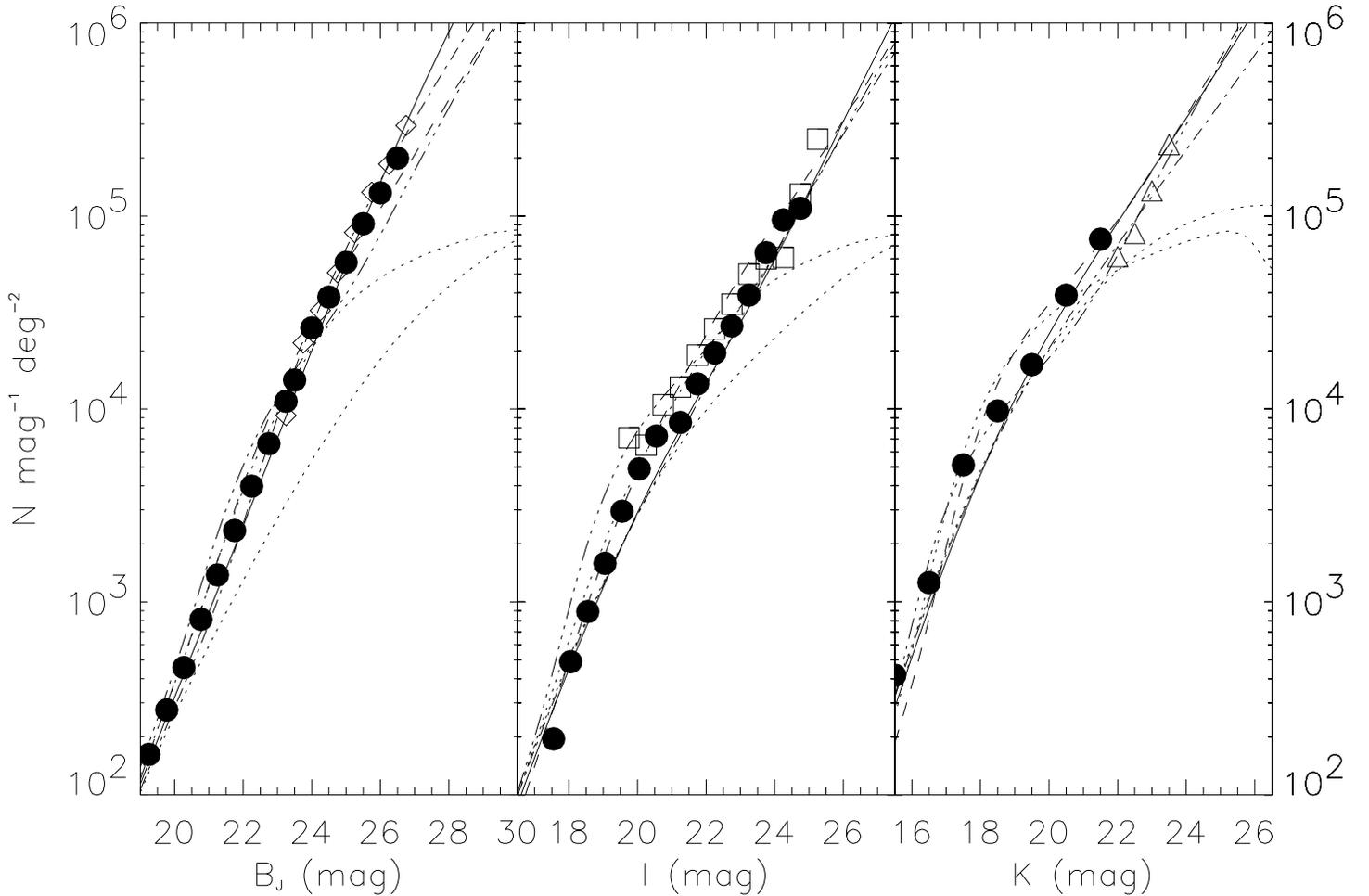}
     \caption{Number vs. magnitude for some models described in the text and
Table~\ref{lumfun}:
dotted lines are for simple NE model in cosmology C (lower) and standard,
passively evolving model (PLE-C, upper). Solid line is model DR2-B,
short dash
BBG-C, dash-dot EDP-C, and dash-triple-dot LZ2-C.  All models are
normalized to fit the observed $B_{J}$ counts at 18.5 mag.
The data plotted, black circles,
is a compilation by McLeod and Rieke (1995) with an addition
of the recent faint $B$-counts from Metcalfe \ea (1995, diamonds),
$I$-counts from Smail \ea (1995, squares), and the faintest $K$-bins
from Djorgovski \ea (1995, triangels)}
     \label{counts}
\end{figure*}

Motivated by some
recent observations and models, I adopt in the following
several cosmological and galaxy-population models, adjust them to fit the
observed number counts and derive their effect in the EBL.
The resulting number count fits are shown in Fig.~\ref{counts}.  I do not
consider redshift distributions in this work; despite considerable
improvements in recent years, the statistics in redshift surveys are
still rather poor compared to magnitude--number counts.

\epsfxsize=18cm

\subsection{Standard model}
\label{stand1}

\begin{figure}
     \epsfbox{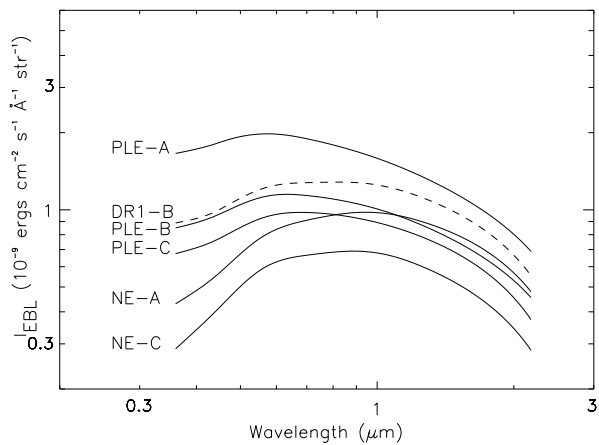}
     \caption{$I_{\rm EBL}$ as a function of wavelength for non-evolving
models (NE-solid curves)
with in flat $\lambda$ dominated universe (A) and high-density universe (C).
Curves labelled PLE  correspond to  models with pure
luminosity evolution (B is for an open, low-density cosmology).
Dashed line is a non-evolving
dwarf-dominated case DR1 (Sec.~\ref{driver})}
     \label{stand1_fig}
\end{figure}

For the sake of comparison, and to examine the effects of
cosmological geometry, I first calculate the EBL from a standard
non-evolving (NE) model.
It is just this type of model
which leads to the excess problem --
the predicted number counts are lower than the observations
by a factor of 4--10
in the $B$-band at apparent magnitudes 24--27 (see Fig.~\ref{counts}).
In Figure~\ref{stand1_fig} are shown the $I_{\rm EBL}$ emerging from NE models
for two different
cosmologies A and C (see Table~\ref{cosmo});
$I_{\rm EBL}$ with cosmology B would lie between these curves.
The $I_{\rm EBL}$ increases by a factor of $\sim 1.5$ when changing
$\Omega_{0}$ from 1 to 0.1 and  introducing a cosmological constant
with a value of $\lambda=0.9$.
The models with large $\lambda$ are found to fit the blue galaxy counts better
(Fukugita \ea 1990; Yoshii 1993), but to fit the counts also in the NIR
requires other effects, \eg including selection effects (Yoshii 1995).

\epsfxsize=9cm

Non-evolving models are clearly unphysical, and models which include
luminosity evolution based on the history of star formation in galaxies,
stellar evolution and population synthesis, should be viewed as the
standard models.
As an example,
the results of a model developed by Arimoto \& Yoshii (1986, 1987
-- AY86-87) are adopted
with the modified
evolution of the UV spectral energy distribution for ellipticals
explained in Yoshii \& Peterson (1991), and for spirals as given in
Arimoto \ea (1992) (AYT92; see their Table 3 for the adopted
evolution).
Typically the evolution in $B$
band for early-type galaxies by $z \sim 1$ is around 2.5 mag (in C cosmology)
and in $K$ around 0.5 mag.

The resulting $I_{\rm EBL}$ with pure
luminosity evolution (PLE) according to the S1-model (see AYT92 for details)
with all cosmologies A, B, and C
cosmologies are plotted in Figure~\ref{stand1_fig}.
This model treats the galaxy as
a ``closed box'', with a typical initial mass function and a star formation
rate; it is similar
to models of Bruzual (1983) and Guiderdoni and Rocca-Volmerange (1987).
(For details see the original papers.)

After testing the other models in AYT92, it was seen that S1 gives
the largest increse of $I_{\rm EBL}$; this increase is by a factor of
$\sim 1.5$ in the optical and NIR and by a
factor of 2 to 4 in the UV compared to the corresponding
no-evolution model.  Note that in contrast to the EBL values of
Yoshii \& Takahara (1988, their Fig.~12), here the evolution also
changes the spectrum --
this is due to underestimated blue evolution inherent in
the AY models,
which is corrected in Yoshii \& Peterson (1991) and AYT92.
Note that the $\lambda=0.9$ cases should be treated
with caution, since the evolution
models themselves are constructed {\em without} considering
the $\lambda$-term; see Martel (1994).

\begin{figure}
     \epsfbox{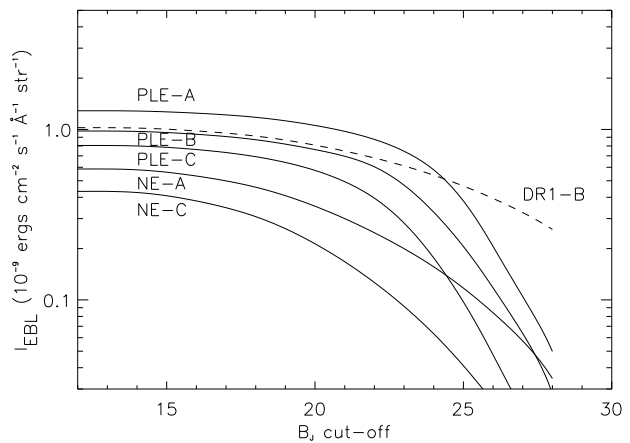}
     \caption{$I_{\rm EBL}$ as a function of $B_{J}$ cut-off magnitude
for the same models as in Fig.~\ref{stand1_fig}}
     \label{stand2_fig}
\end{figure}

The effect of the cut-off magnitude is clearly seen in
Fig.~\ref{stand2_fig} where exactly the same models are plotted as in
Fig.~\ref{stand1_fig} but as a function of
$B_{J}$ cut-off magnitude.
The contribution of galaxies
brighter than about $B = 15$
mag is negligible; however, a measurement
of EBL which cuts off objects brighter than \eg 20 mag, produces a value
which is about half of the total.  Thus,
when reporting either a measured or
a theoretical value of EBL it is important to specify the limiting
magnitude used -- it is not always clear in the literature, whether a
``total'' EBL value is used, including the brightest magnitudes, or a value
corresponding to
some cut-off magnitude.  In all the figures showing the SED of the EBL
in this paper, the cut-off magnitude is chosen to be sufficiently bright,
that the values reflect the output of all galaxies.

The amount of
EBL coming from fainter sources depends critically
on the properties of galaxies.  As an example, a dwarf-dominated model DR1
(see next section) is also plotted in Figs.~\ref{stand1_fig}
and~\ref{stand2_fig}.  It has the same spectral shape and overall normalization
in the UV and optical bands as the evolving PLE-B,
but the light is coming from a separate class of much fainter galaxies;
see Sects.~\ref{nonevdwarfs} and~\ref{fractions} below.

\epsfxsize=9cm

The results in AYT92 and AY86-87 are given for a
galactic age of 15 Gyr.
Naturally the age of the universe and the age of the galaxy
can be less (or more) than this depending on
cosmology and the choice of $z_{\rm for}$.
Following YT88
an approximation is used, where the present-day magnitude is simply
taken to be the value at a galaxy age of $t_{0} - t_{z_{\rm for}}$;
the error is largest for
the $\Omega_{0},\lambda_{0} = (1.0,0.0)$ cosmology, but is not
significant, as the magnitudes do not change much after 10 Gyr.

\subsection{Dwarf-dominated models}
\label{driver}

$K$-band counts do not show nearly so large an excess over predicted
numbers from no-evolution models,
(\eg Gardner, Cowie, Wainscoat, 1993); thus a common
practice has been to add a new {\em faint blue}
galaxy population to fit both the $B$ and $K$ bands.
This procedure fits well
the results of the HST Medium Deep Survey, which seem to
establish that
the faint excess is due to predominantly faint and blue
galaxies with irregular morphology.

In addition, a recent local LF determination by
Marzke \ea (1994; see also discussion and references therein)
shows a significantly steeper faint end of the LF for dwarf
galaxies than would be expected from the ``standard'',
overall Schechter LF adopted for this work.

In the next section I present a couple of variants of dwarf-dominated models.

\subsubsection{Non-evolving dwarfs}
\label{nonevdwarfs}

\begin{figure}
     \epsfbox{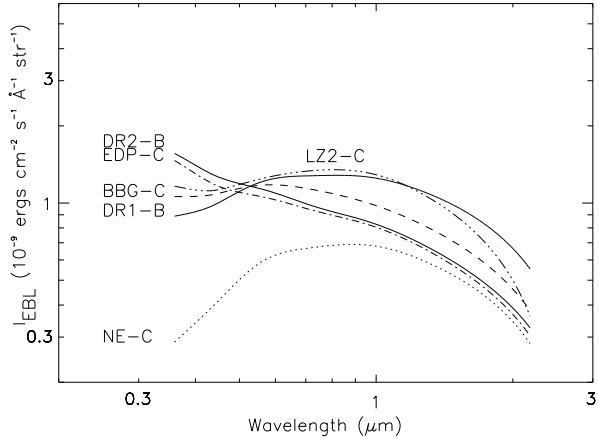}
     \caption{$I_{\rm EBL}$ as a function of wavelength for several
dwarf-dominated
models.  See Table~\ref{lumfun} for characteristics of plotted models.
NE-C is shown for comparison}
     \label{dwarf1}
\end{figure}

Driver \ea (1994) find
good fits to number counts by adopting a steep ($\alpha=-1.8$)
faint-end slope of
the LF for two separate extra populations of dwarf galaxies (dI and dE;
see Table~\ref{lumfun}, model DR1).
The calculated $I_{\rm EBL}$ is shown both in
Fig.~\ref{stand1} and Fig.~\ref{dwarf1}.
The expected $I_{\rm EBL}$ is nearly 3
times larger than the reference NE model (cosmology B) in the blue
wavelengths and less than 2 times
larger in IR.  The $I_{\rm EBL}$ in this case (note that there
is no luminosity evolution) is at the same level as for passive luminosity
evolution in the same cosmological model.

\epsfxsize=9cm

The effect is especially dramatic at faint magnitudes,
because of the high abundance of dwarfs; this can be seen from
Fig.~\ref{stand2_fig}, where DR1 is plotted
as a function of the $B_{J}$ cut-off.
The EBL coming from sources fainter
than 25 mag is nearly 10 times greater in the dwarf-dominated model
than in a corresponding NE model.

\begin{figure}
     \epsfbox{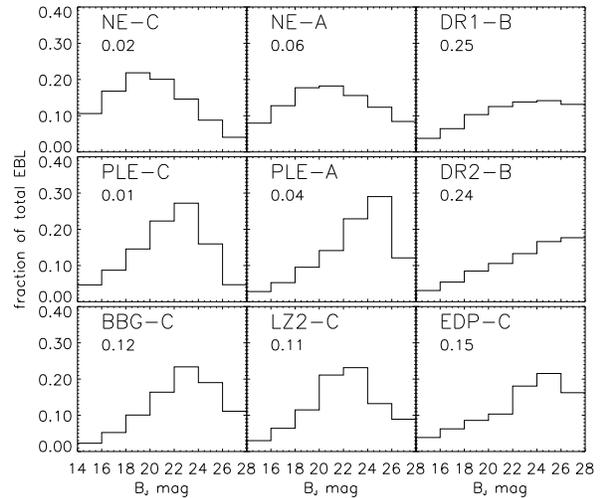}
     \caption{The fraction of total $I_{\rm EBL}$ coming
from magnitude intervals
ranging from 14--16 to 26--28 mag.  The
number printed in each histogram is the fraction of total $I_{\rm EBL}$
coming from fainter than $B = 28$ mag}
     \label{magbin1}
\end{figure}

The same effect can be seen in Fig.~\ref{magbin1}, where
the ratio of EBL contribution from a given magnitude interval
to the total EBL is plotted.
One can see that in the DR1 model
the contribution coming from magnitude intervals keeps on rising up to the
faintest
magnitude ranges, and $\sim 25$ \% of total EBL is still
beyond the reach of current deep galaxy surveys.

{}From Fig.~\ref{magbin1} one can also deduce
the qualitative effects of cosmology and
luminosity evolution:
A non-zero cosmological constant, besides increasing the overall $I_{\rm EBL}$,
pushes the magnitude range where most of the
EBL originates fainter by some
2 -- 4 magnitudes (compare NE-C to NE-A and PLE-C to PLE-A) -- a
natural effect of increasing the volume and age of the universe.
Including pure luminosity evolution,
and $z_{\rm for}=5$, increases the $I_{\rm EBL}$ especially in
the range $B=22-24$ (model PLE; also BBG and EDP explained
in Sect.~\ref{evdwarf} below).
If the formation of galaxies takes place closer to the present epoch
($z_{\rm for}=2$ in LZ2, Sect.~\ref{recentgal}),
the corresponding enhanced magnitude range is brighter.

An even better overall fit to number
counts in the bands up to $K$ is found if the DR1
model is slightly modified (labelled DR2):
instead of dE and dI galaxies, only
dI's are used (with $\alpha = -1.6$),
and their SED is assumed to be flat in $f_{\nu}$ and hence
very blue ($f_{\nu}$=const. i.e. $f_{\lambda} \propto
1/\lambda^{2}$).  The resulting fit is in Figure~\ref{counts} and
$I_{\rm EBL}$ for this model is
in Figure~\ref{dwarf1}.  Now the UV and blue luminosities are nearly
doubled from the previous dwarf-model.

\epsfxsize=9cm

\subsubsection{Dwarfs and evolution}
\label{evdwarf}

While not considering redshift distributions in detail in this work,
the previous models do, however, produce a sizeable
low-redshift peak which is hard to reconcile with observations
(\eg Glazebrook \ea 1995).  A way out would be
to let the dwarf population evolve in some way.
Treyer and Silk (1994; see also Cole, Treyer and Silk, 1992)
introduced a model for the evolution of $B$- and
$K$-band LFs, in which the slope of the faint end steepens and $L^{\ast}$
increases with increasing redshift.
With this idea in mind, we let the slope of the new
evolving dwarf population (EDP)
steepen rapidly until $z=0.25$ and remain constant at higher redshifts.
Recent observational evidence for a steepening LF comes from Ellis \ea (1996).
The characteristic magnitude brightens from $-18.2$ to
$-20.5$ ($H_{0}=50$) by redshift $z=1.2$ and then drops again, mimicking a sort
of star-burst.

The resulting $I_{\rm EBL}$ is plotted in Fig.~\ref{dwarf1};
it shows the characteristics
of a flat-spectrum dominated population as in the DR2 case above.
The success in fitting the number counts comes from these {\em very}
blue objects.  In the infrared region the $I_{\rm EBL}$ does not differ
significantly from a NE model.
The UV light is some 5--6 times greater compared to the NE
model and twice or more greater than in the non-flat-spectrum
dwarf-dominated models.

In the redshift distribution this model would give many more high redshift
detections than the previous models.
$I_{\rm EBL}$ coming from different magnitude ranges is shown
in Fig.~\ref{magbin1}.  About 30 \% of EBL
comes from sources fainter than $B=26$ mag.  In the
DR2 models the corresponding value is 40 \% ; in addition to different
redshift distributions the EDP model shows an earlier drop in the
$I_{\rm EBL}$ contributing to faint magnitude ranges.

Introducing standard luminosity evolution into the giant population obviously
changes
the needed properties of the dwarf-population.  The following models have
passive luminosity evolution included in the giant population, which results
in an enhancement of the EBL as described in Sect.~\ref{stand1}.

The Canada-France Redshift Survey (Lilly~et~al.\ 1995a)
has recently provided a picture of the evolution of the galaxy LF out to
$z \sim 1$.  A
galaxy population model which qualitatively reflects the  findings
of Lilly \ea (1995b) is constructed:
i) little change in red galaxies -- the model here has
the standard passive evolution; ii) extra brightening of the
blue galaxies (model labeled BBG); iii) excess of faint galaxies --
an additional, non-evolving
blue population of galaxies is included.  See Table~\ref{lumfun}
for details.

The fit to the counts is in Fig.~\ref{counts} and the
resulting $I_{\rm EBL}$ is
plotted in Fig.~\ref{dwarf1}.  As expected, the blue luminosity has increased
compared to DR1 due to evolution in the blue population.  The numbers of the
giant population as a whole
are slightly overpredicted in the range
$I=19-22$ as compared to the MDS results (see Sect.~\ref{fractions}).
To fit these better the strength of the
luminosity evolution of ellipticals could be somewhat decreased
(the spectral shape of the EBL would become bluer) which would
still be in accordance with Lilly \ea (1995b).  The first direct measurement
(Im \ea 1996; HST MDS data) of the luminosity evolution of
ellipticals also suggests this: while ellipticals are seen to evolve, the
evolution is rather mild, about $0.5 \sim 1$ mag by  $z\sim 1$ in $I$-band.

All of the above dwarf-dominated
models give an $I_{\rm EBL}$ around 1 \cggs over the optical wavelengths.

In faint galaxy -dominated models,
the choice made in Sect.~\ref{mod}, to integrate
to 8 mag fainter than $M_{B_{J}}^{\ast}$, could in principle
significantly affect the results.  Taking a
severe case, DR2 (see numbers reflecting the 'unseen' population in
Fig.~\ref{magbin1}), it is found that the $I_{\rm EBL}$ decreases
by $\sim 20$\% if
the integration goes only to 4 mag below $M_{B}^{\ast}$, and increases by
$\sim 15$\% if there is no cut-off at all in $M_{B_{J}}^{\ast}$.
The percentages are slightly larger in the blue than in the IR.  It was also
found that the difference arises almost entirely because of the inclusion or
exclusion of galaxies {\em fainter} than those present
in any existing number counts.

\subsubsection{Recent galaxy formation}
\label{recentgal}

Cowie \ea (1995) have recently provided observational evidence of strong
ongoing
star formation at relatively low redshifts.  Adopting passive luminosity
evolution and using $z_{\rm for}=2$ for all giants, it is
impossible to fit the number counts.
All cosmological models have a bump around B$\sim 22$ mag, the most
severe case being the high $\lambda$ model, and
all still have a deficit of faint galaxies, especially the
high $\Omega$  model.

In an attempt to force a low $z_{\rm for}$
to fit the counts, I construct a new model (LZ2).
To have the smallest blue bump, cosmology C is chosen,
$z_{\rm for}$ for all giant galaxies, and
an extra dwarf population is added to account for the faint counts.
Here a different approach is taken to qualitatively avoid
the excess low redshift peak -- the extra blue dwarf population, which has
$\alpha = -1.8$, {\em disappears} at low $z$.
See Table~\ref{lumfun} for details.

The fit can be seen in Fig.~\ref{counts}.  Choosing a low-density universe
would reduce the need for the faint galaxies, but it would also  require
some additional effects, \eg
selection or dust effects, to remove the blue bump.
Campos \& Shanks (1995) have recently shown
that by including a simple dust model one can get rid of
a similar bump in the number counts.

The resulting $I_{\rm EBL}$ of model LZ2 is shown in
Figure~\ref{dwarf1}. It
shows an excess of light shortward of $1 \mu$m compared to other models.
This is natural, as the first bright stages of
star formation are more recent and the bright UV and
blue radiation from this era
is redshifted to visual and red bands. If there is significant star formation
going on at very low redshifts, it would show in an even bluer SED of EBL.

If the  formation time of galaxies is pushed further back in time, say,
to $z_{\rm for} \geq 10$, the $I_{\rm EBL}$ does not change significantly in
any cosmology.
The change adds more time for the EBL to emerge, but sources are extremely
faint and thus contribute little; in the case of luminosity
evolution $I_{\rm EBL}$ would be somewhat
lower, because the bright early stages of the early type galaxies are further
away.

\subsubsection{Models with merging galaxies}

Some proposed explanations to the faint excess and redshift distributions
include merging scenarios
which preserve the luminosity density of galaxies $(\varphi^{\ast}_{z}
L^{\ast}_{z}=$const); \eg Rocca-Volmerange \& Guiderdoni (1990) and Broadhurst
\ea (1992).   In general, it is difficult to fit both the blue and
IR data with these models, without some additional ingredients (like extra
dwarf populations).  Also, because of the conserved luminosity density
the $I_{\rm EBL}$ is not much different from
the NE case; therefore, these are not further investigated here.

\section{Low surface brightness galaxies and EBL}
\label{lsbg}

\subsection{Detecting galaxies}

So far the galaxies have been treated as  point sources with total
magnitudes.  However, real galaxies
are generally extended objects and
are selected by their surface brightness.  A commonly used form for the
surface brightness profiles $g(r)$ of galaxies is
\begin{equation}
\label{surf}
g(\beta) = \exp(-a_{n}(\beta)^{1/n}) \ \  ;\beta \equiv r/r_{e}.
\end{equation}
For ellipticals and
the bulges of spirals $n=4$, and for spiral disks
$n=1$; $r_{e}$ is the effective radius of a galaxy
within which half of the total luminosity is encircled and thus $a_{4}=7.67$
and $a_{1}=1.68$.  This profile is further convolved with the point-spread
function of the telescope and atmospheric seeing when observing the galaxies.

Yoshii (1993) has considered faint galaxy counts, taking into account the
detection and selection effects.  He has used essentially
the same galaxy models as adopted in this paper for the NE and PLE
mo\-dels.  Taking the selection effects into account, the most radical
effect is the sharp drop of the number count slope after a certain
magnitude (see his Fig.~5.).
This drop is consistent with Tyson's (1988)
$B_{J}$ observations.  Note
that this drop is caused by observational effects, and that the {\em
intrinsic} count slope, which affects the EBL, can still be rising.
Yoshii finds that a low-density university, with a non-zero cosmological
constant and standard luminosity evolution,
gives the best fit to the galaxy counts in different bands, allthough the
model still predicts a slight deficit of galaxies in the $B$-band
in the range 22--26 mag.  Addressing specifically the $K$-band counts Yoshii
and Peterson (1995) conclude that it is not possible to rule out
the nonzero $\lambda$ models using the deepest $K$-band data.
In the context of this work Yoshii's result is interpreted as stating that
the galaxy population of
model PLE-A, giving the $I_{\rm EBL}$ seen in Fig.~\ref{stand1_fig},
would fit the observed counts, if selection effects would be given due
consideration.

\subsection{Low surface brightness galaxies}
\label{lsbgal}

Yoshii assumed that the Freeman law (Freeman 1970)
holds for spirals, i.e.\ that the central
surface brightness is constant, $\mu_{0} = 21.65 \pm 0.35$ $B_{J}$ mag
arcsec$^{-2}$.
However, a significant number of spirals with
surface brightnesses more than $3\sigma$ away from the Freeman relation
have been found, and the Freeman law
should thus be abandoned,
or at least modified (\eg MG95; de Jong 1996).

FMG95 have considered
the faint galaxy counts including selection effects
in a similar way to Yoshii, but relaxing the Freeman law.
They have concluded that
it is possible to include a large number of LSB galaxies to the LF without
violating the constraints on local field-galaxy luminosity functions.
McGaugh (1994 and 1995) stresses the point that
a population which is observed at intermediate redshifts in deep
surveys is undetectable in shallower galaxy surveys because of
very low surface brightness;  an even more
severe effect than complete
non-detection arises from systematic under-estimation of the fluxes of
LSB galaxies.

FMG95 consider two cases, models A and B, in which different
assumptions
have been made on the dependence between the central-surface-brightness of
a spiral galaxy and its luminosity (see Fig.~1 of their paper):
A) surface-brightness is independent of
luminosity; B) central surface-brightness
decreases with decreasing luminosity for
galaxies fainter than $L^{\ast}$ and is a constant,
defined by the Freeman value for giants ($L > L^{\ast}$).  They also
allow scatter around these relations.  For S0's the Freeman value
is used, and for ellipticals a relation of
$\mu_{e} = 1.20 \log (L/L^{\ast})+21.16$
is adopted, i.e.\ central surface brightness increases with
decreasing luminosity.

Models A and B are two extremes and demonstrate the effect of the
bivariate luminosity function $\Psi(h,\mu_{0})$, where $h$
denotes the scale length of a galaxy,
over the normal
$\Psi(M)$. Model A leads to an increase of the normalization constant
$\varphi^{\ast}$ of the LF
while model B steepens the slope of the faint end of the
luminosity function (see also McGaugh 1994).  Model B seems to be closer to
reality as observations (MG95) point toward
a decreasing $\mu_{0}$ with decreasing luminosity.  But a lower $L$ does not
necessarily imply a {\em dwarf} galaxy -- in fact, the
assumption made in FMG95,
that the size $h$ and $\mu_{0}$ do not correlate, is supported by many
studies, the most recent ones being de Blok \ea (1995) and McGaugh \ea (1995).

FMG95 constructed a simulated
galaxy-catalog which they ``observed'' using a Monte Carlo method,
closely following real
observational criteria.  Using constraints from the {\em observed}
properties of
galaxies (LF very close to the one adopted in Sect.~\ref{lumfun}),
they produce a possible set of the {\em
intrinsic} properties of galaxies.

\begin{figure*}
     \epsfbox{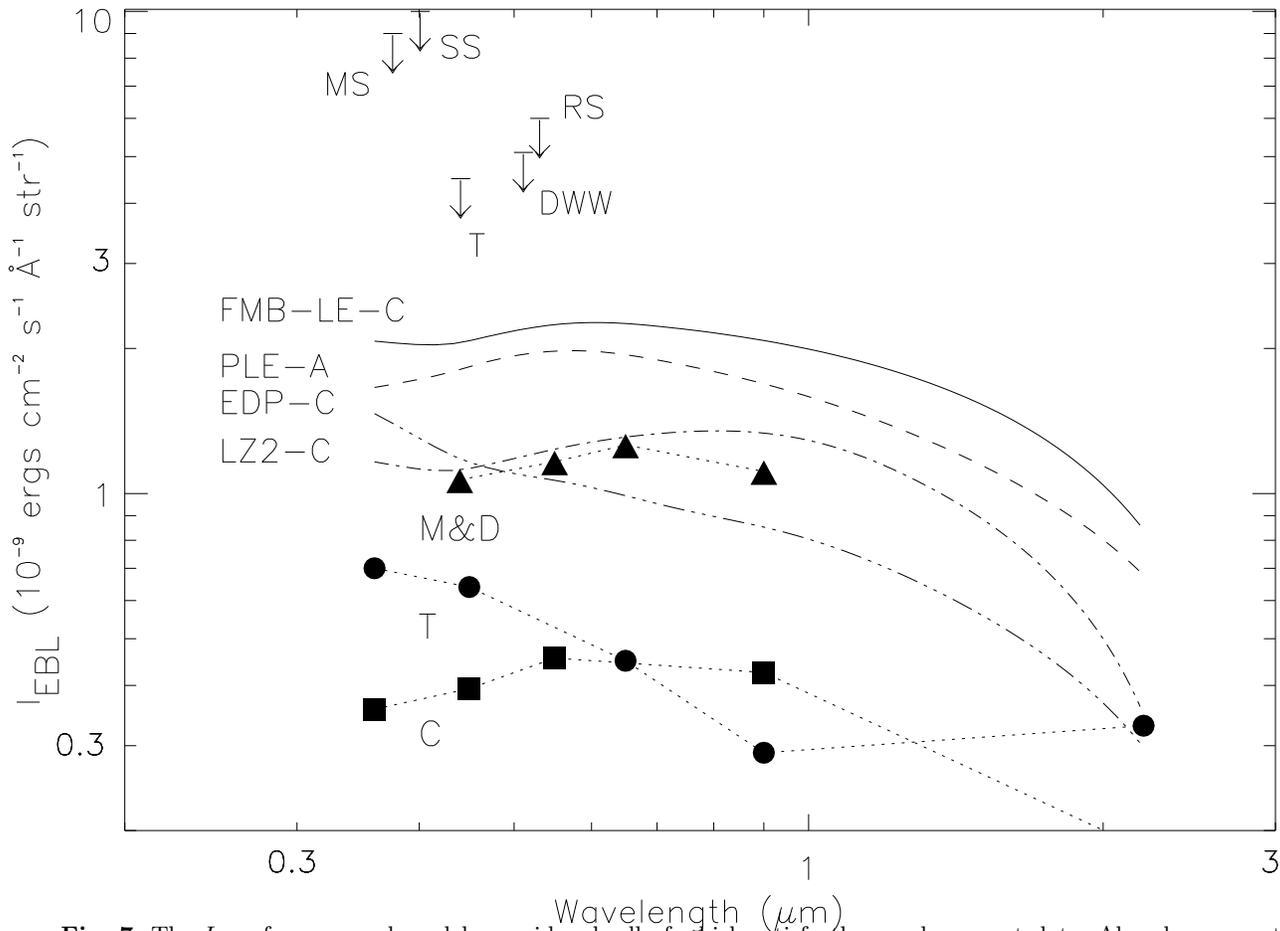}
    \caption{The $I_{\rm EBL}$ from several models considered,
all of which satisfy
the number count data. Also shown are the results of
Tyson (1995, T), Cowie \ea (1995, C), and Morgan \& Driver (1995, M\&D)
-- the last one is an extrapolation of actual observed counts, see text.
Observational upper limits are from Mattila \ea (1991);
see Sect.~\ref{surfphot}}
     \label{select_fig}
\end{figure*}

The number count predictions produced by isophotal selection in their
scenario still somewhat
underestimate the observed counts.  The model, especially in the $B$-band,
would come closer to the data if an open universe (cosmology B) is assumed.
Also,
FMG95 note that the number count predictions leave room for regular luminosity
evolution: the $I_{\rm EBL}$ of a model with the LF
properties and SED's from Table 1 of FMG95 is included
in Figure~\ref{select_fig}
-- but with the passive luminosity
evolution added (FMB-LE, C cosmology).
It exhibits a dramatic rise of $I_{\rm EBL}$;
the overall $I_{\rm EBL}$ is 4--7 times brighter
than in the reference NE model.  The rise is due to: i) a 2.3
times higher normalization of the LF as compared to the NE model;
ii) luminosity evolution brings in
an enhancement by a factor of about 2 in blue and 1.5 in NIR as described in
Sect.~\ref{stand1}; iii) about a
factor of 1.5 rise in the blue wavelengths (dropping to a factor of 1.1 by
$K$-band) is due to a steep faint end slope
($\alpha = -1.8$) of Sdm-class of galaxies, which
constitute a quarter of all galaxies in the FMB model.
The value of EBL is actually rather sensitive to a steep faint end slope --
using $\alpha = -2.0$ for the Sdm's of this model,
would result in another enhancement by a factor of 1.5.

Almost the same amount of $I_{\rm EBL}$ would come
out of a FMB model with no luminosity evolution but with a B cosmology.
Figures as a function of magnitude limits are not presented here since
they require an explicit treatment of the isophotal magnitudes and
selection effects --  in the figures which show the
SED of EBL it does not matter
because the $I_{\rm EBL}$ is the total integrated flux, whether
the galaxy is detectable or not.

The preceding model FMG-LE has to be
taken with caution, as evolution was not
included in FMG95's simulation.  But any significant over-estimation of the
counts should not be expected:  As a comparison McLeod and Rieke (1995)
include
in their similar investigation of
faint galaxy models a LSB population of the same size as all giant galaxies
and properties close to the Sdm class
(except, of course, the surface brightness), and
the expected counts fit or still fall below the observations.

\epsfxsize=17cm

\section{Discussion}

\subsection{Comparison with EBL values derived from specific counts}

The pure luminosity evolution model
PLE (cosmology A), EDP (C), LZ2 (C), and FMB-LE (C) are plotted
in Fig.~\ref{select_fig} as
representatives of models which agree
with the number counts in the blue and IR bands.
It is interesting to compare these predictions with
the EBL integrated from
number count results by Tyson (see Fig.~12 of Tyson 1995),
by Cowie \ea (1994), and those done by the
Cardiff group (Driver \ea 1994; Morgan \& Driver 1995).

The spectral distribution of EBL from
Tyson's (1988, 1995) faint
galaxy counts is significantly blue.
Of the models
discussed in this paper, the DR2 and EDP models, which include a flat spectrum
population, show the same shape in UV to optical wavelengths.
The difference in the
level of EBL is mainly due to Tyson's {\em flattening}
counts beyond $\sim 27$ $B_{J}$ mag, an effect
the DR2 and EDP models do not have;
also a correction term of some 20\% for
an undetected population (fainter than $B_{J} \sim 27$ mag)
could be justified on the basis of
dwarf-dominated models.  A more
significant correction could arise from
very low surface brightness galaxies;
though Tyson (1995) argues against this,
the models of FMG95 and McLeod and Rieke (1995)
do show that such a population could be accommodated in
the local universe without violating the observed counts and LFs.

Cowie \ea (1994) have integrated the light emerging from
galaxies in the $K$-selected counts up to $K=22$ mag, corresponding
roughly to
$B=25.5$, and obtained a
value of $I_{\rm EBL} \approx 0.4$ at blue wavelengths in our units.

It is seen that
the shape of the spectral distribution of Cowie's EBL agrees very well with
the dwarf-dominated models presented here (DR1, BBG, LZ2).
The level of $I_{\rm EBL}$ is lower, but looking at Fig.~\ref{magbin1},
especially the model DR1,
it is clear that the {\em total} $I_{\rm EBL}$ from galaxies
could well be twice the given value.  In that case, the level of Cowie's
EBL comes close to the models.  Of course the correction factor must
depend on the wavelength because of the different colors of galaxies --
actually Cowie \ea (1994) do
show (their Fig.~13) that the SED of EBL is clearly
bluer the fainter magnitude bin one takes.  So, extrapolating to much fainter
magnitudes the IR EBL would rise only slightly (in fact, doing the same
calculation as in Fig.~\ref{magbin1} one notices that the bulk of EBL
in $K$-band comes from galaxies around $K \sim 17$), while the blue light would
rise more rapidly.  Thus also Cowie's SED of the EBL might well come close
to the DR2 and EDP models.

That the $K$-band EBL is close to predicted values is understandable
because the light there is coming from very different star-populations than
at blue wavelenghts; see Sect.~\ref{fractions} below.
The issue of faint galaxies is almost
separated from the EBL in the IR; even in the
case of extreme dwarf-domination, clearly over half of the total light comes
from ordinary giant galaxies.

Morgan \& Driver (1995) present values of EBL derived from number counts in the
$B$, $V$, $R$, and $I$ bands by the 'Hitchhiker
camera' on William Herschel Telescope.  They also extrapolate their results
according to a dwarf-dominated galaxy model in the same way as above (see
Figs.~\ref{magbin1} and~\ref{stand2_fig}).  These extrapolated
values of $I_{\rm EBL}$ are shown in
Fig.~\ref{select_fig} -- they are essentially the same as those predicted by
DR1 of this paper since the assumed underlying galaxy populations are almost
identical.

All in all, the current galaxy counts suggest a much bluer EBL spectrum
than predicted from standard flat LF's of giant galaxies.

\subsection{Comparison with the EBL upper limits from surface photometry}
\label{surfphot}

The previous comparisons essentially compare different galaxy models to
observed  galaxiy counts.  However,
a whole new area is opened up when comparing the models to the
measured EBL, possibly including  diffuse radiation in addition
to light from galaxies.

The value of the measured EBL is still an
unsettled matter.
The model results of previous sections are compared
with the limits obtained so
far.  Mattila (1990) and Mattila \ea (1991, MLS) have
rewieved the status of EBL determination from surface
photometry.  So far, only upper limits are available in the optical, UV
and near-IR bands.  Some of these upper limits are plotted in
Figure~\ref{select_fig}.  In several cases, as discussed by MLS,
the upper limits originally given by the
author(s) have been too stringent due to insufficent consideration of
the atmospheric or photometric sources of error.  The upper limits shown
if Fig.~\ref{select_fig} are in each case the more conservative ones
according to Table 2. of MLS -- Dube \ea (1979) DWW,
Mattila (1990) MS, Roach \& Smith (1968) RS, Spinrad \& Stone (1978) SS, and
Toller (1983) T.
These upper limits are still at least a factor of $\sim
5$ higher than the EBL predicted by most of the models and a factor of
$\sim 3$ higher than the models involving LSB galaxies with selection effects
taken into account.

\subsection{Uncertainty in the UV-spectra of ellipticals}
\label{uv}

Following Yoshii \& Peterson (1991) the effect of
uncertainty in UV-spectra of ellipticals is tested.
Calculating the NE model as
in Sect.~\ref{stand1}, but using the SED of NGC 4649, a ``UV-hot''
case for ellipticals and S0 galaxies, it was found
that the effect on the $I_{\rm EBL}$
is negligible.  When evolution is included, the shape of the UV-spectrum of
ellipticals is made more important because of their bright
early stages: in blue bands $I_{\rm EBL}$  increases by a factor of
1.2 over the  evolution model with a regular SED.

If the Yoshii \& Peterson's $x$-term concerning UV-evolution
(cf.\ their Eq.(1)),
is changed from 0.2 to 1.0
(this simulates Bruzual's (1983) evolution model for UV-hot galaxies;
it is rather extreme to assume this for {\em all} ellipticals),
$I_{\rm EBL}$  increases by a factor of $\sim 1.8$ at optical wavelengths.
However, if also the UV-hot SED mentioned above
is adopted, the increase of $I_{\rm EBL}$ is large,
by a factor of 4 to 6 in optical bands compared to PLE-C.
These values are already at the upper limits
from surface photometry (cf.\ previous section).

The UV-hot models result in an
excess bump over the observed $B_{J}$ counts
in the magnitude range 20--24.  This effect is similar to
that found by Guiderdoni \& Rocca-Volmerange (1990) for UV-hot evolution
models, which are only
consistent with number count data in the case of
large  galaxy formation redshift ($z_{\rm for} \ge 10$).

As mentioned in Sect.~\ref{mod},
the galaxies are assumed to be opaque to Lyman
continuum photons.  In reality the gas in galaxies is clumpy and thus
some of the these photons could escape.  To examine the size of this effect,
the models were calculated also {\em totally without}
the above cut-off.  In the
$U$-band, in models with luminosity evolution of galaxies, the EBL would
rise by about 24, 14, and 7 \% with cosmologies A, B, and C, respectively.
In $B$
band the difference is smaller, a $\sim 9$ \% rise of the EBL in cosmology A
and about half of this in B.
If no luminosity evolution is included, or if $z_{\rm for}$ less than
$\sim 4$, the effect on the SED of the EBL is negligible.

\subsection{Contributions of different galaxy types}
\label{fractions}

Any reasonable changes in the fraction  of giant galaxy types do not affect
the $I_{\rm EBL}$ significantly.  However, it is
interesting to determine from which
population(s) the EBL mainly emerges both in standard and dwarf-dominated
models.

\begin{figure}
     \epsfbox{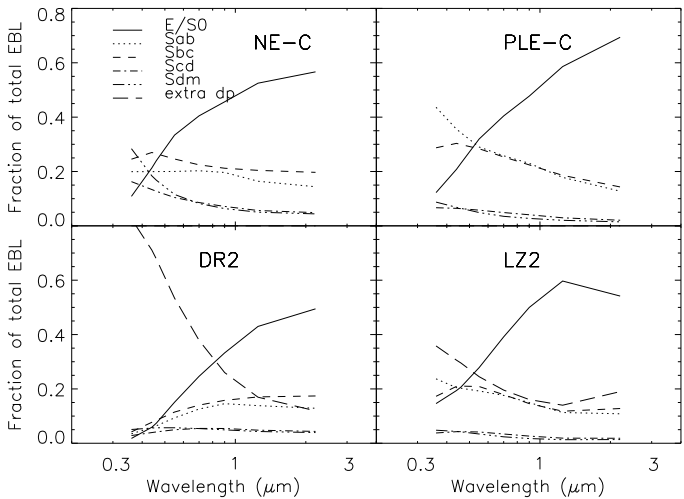}
     \caption{Fractional contribution of different galaxy types to total
EBL in several models considered}
     \label{frac}
\end{figure}

Fig.~\ref{frac} shows the contributions of different
galaxy types to $I_{\rm EBL}$.  As a reference, the upper right figure
has the contributions
of galaxies in
a regular, no-evolution model -- the upper right one has an added
passive luminosity
evolution, which makes the earlier types more dominant (the evolution
actually dims the late types, while considerably brightening the E/S0's).
In the near-infrared, by far most of the EBL is expected to come from
ellipticals.

As an example of the dwarf-dominated models, the DR2 model
shows the dominance of
dI-dwarfs, whereas the the population of blue dwarfs combined with PLE for
giants (LZ2) dominates only in the UV.

\epsfxsize=9cm

There is an interesting implication if a high (higher \eg than YT's
predictions) EBL value results from a large population of dwarfs or
LSB galaxies -- the overall fluctuations of the EBL
would be expected to be {\em lower}.  This is because the extra population
is most probably more evenly distributed:
Mo \ea (1994)
show that LSB galaxies are less strongly clustered than giant galaxies and
this seems to be
true also for faint blue galaxies (\eg Tyson 1995), which might well
be dwarfs, as stated earlier.

Thus, the result of Sasseen \ea (1995), which
indicates that the early measurement of EBL fluctuations by Shectman
(1973, 1974) gives too high a result due to contamination by galactic cirrus,
actually leaves room for a {\em larger} value of total EBL.  Also, when Davies
\ea (1994) consider the fundamental limit set out by the EBL fluctuations
on the detection of galaxies with very low surface brightness, the situation
with a 'high' value of EBL might not be as difficult as they suggest.

\begin{figure}
     \epsfbox{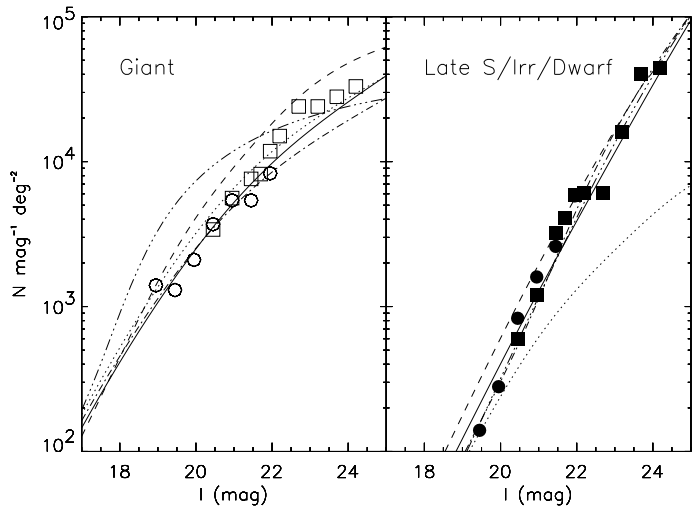}
     \caption{The morphological number counts from the HST Medium Deep Survey
compared to some models presented in this work.  The circles are from
Glazebrook \ea (1995) and the squares from Driver \ea (1995).  The magnitudes
have been corrected by 0.2 to move
from the I814 band to Johnson $I$ (Fukugita \ea 1995).
The dotted line is the simple NE-C
model, which shows a clear underestimation of dwarf counts,
but at the same time
fits the giant counts very well.
The solid line is model DR2, short dash
BBG, dash-dot EDP, and dash-triple-dot LZ2, all of which fit the dwarf
counts remarkably well}
     \label{morpho}
\end{figure}

While I have not
specifically tried to fit the morphologically-split
number vs.\ magnitude counts,
the predictions of models presented in this work are compared to
HST Medium Deep Survey results in Fig.~\ref{morpho}.
The data shown are  the counts of
Glazebrook \ea (1995) and Driver \ea (1995b)
with the giant galaxies (ellipticals and spirals grouped
together) and dwarfs plotted separately.
Glazebrook \ea (1995) have labeled the latter group as
``irregular and merging'' and Driver \ea (1995) as ``late type spirals''.
The model curves from the present work include
the Sdm-class and any extra dwarf galaxy population for this group.

All the models with dwarf populations, invoked to fit the
total number counts, also fit the Late type/Irregular/Dwarf -morphological
counts remarkably well, even though the
dwarf populations have somewhat different characteristics.
The models with a non-evolving giant population (in NE, DR2, and EDP) seem to
fit the giant data better than the ones with luminosity evolution
(BBG and LZ2, short-dashes and dash-triple-dot respectively), the best perhaps
being the NE-C model.  For estimates of
uncertainties in the data see Glazebrook \ea (1995) and Driver \ea (1995b).

\epsfxsize=9cm

\section{Conclusions}

The main results of the work can be summarized as follows:

(1) Models which predict number counts that are consistent with
observations can
have clearly different $I_{\rm EBL}$ levels and spectral shapes.

(2) An EBL value of $I_{\rm EBL} \sim 1 \cdot$ \cggs appears to be
of the correct order of magnitude.
All the models considered without specific selection effects give
a value at or just above this level in the optical and dropping off to
$\sim 0.4$ by $K$-band.  These could be divided roughly into two,
the first group having a flat SED of EBL (in $f_{\lambda}$) in the UV to blue
and the second group having an enhanced UV EBL, up to 2 in the above units.
Models with the $I_{\rm EBL}$ dropping towards the UV
(\eg Yoshii \& Takahara 1988) --
which would constitute a'third group' in the above distinction -- are
found to be inconsistent with observed galaxy  counts.

(3) The distribution of EBL in different
magnitude ranges exhibits a strong dependance of cosmology,
galaxy population, and evolutionary model;
universes producing same $I_{\rm EBL}$ with different galaxy
populations can in principle be seperated using both the SED of EBL and its
intensity as a function of cut-off magnitude.

Most importantly, different models predict very different EBL levels
beyond the current (and some future) magnitude limits.

(4) The situation changes
when selection effects due to isophotal
detection, low surface brightness effects, {\em and} LSB galaxies are included.
In principle the
LSB galaxies could help to produce a very high-intensity EBL.
Even considering more realistic ideas about the properties of LSB's, one
can still produce an $I_{\rm EBL}$ of about $2 - 3 \cdot$ \cggs.
It is with the LSB models
that the present upper limits (around 5--9 in the same units)
of the observed EBL start providing constraints.

It is sobering to see how much the  large uncertainties in the
surface brightness
characteristics of galaxies can affect the EBL and the galaxy  counts.

If all galaxies could be seen, then the EBL and galaxy counts
would not give independent results (apart from non-galactic contributions
to the EBL; see below).
However, especially if there is a large population of LSB galaxies, the
EBL -- as a function of wavelength and cut-off magnitude --
provides a powerful tool for observational cosmology to complement
galaxy counts and redshift distributions.

Furthermore, assuming that we have a measured value for the EBL and that
the galaxy model predictions provide an accurate prediction, then the
difference would account for any previously unknown sources of radiation,
such as decaying particles or any radiation of
intergalactic or/and pregalactic origin.

\begin{acknowledgements}
I wish to thank Kalevi Mattila for initiating the
whole work and for offering good advice
along the way, and Stacy McGaugh, the referee, and John Huchra for
helpful comments, and finally Eric Woods for thoroughly checking the
language.  This work was partially supported by the Finnish Academy of Sciences
(V\"ais\"al\"an s\"a\"ati\"o) and SAO predoctoral fellowship.
\end{acknowledgements}

\appendix
\onecolumn
\section{Cosmological ingredients}
\label{app}

The expansion rate of the universe is determined by the mass density $\rho$,
cosmological constant $\lambda$ and space curvature $k$.
Non-dimensional parameters $\Omega = 8 \pi G \rho / (3H^{2})$,
$\lambda = \Lambda c^{2} / (3H^{2})$, and $\kappa = kc^{2}/(a^{2}H^{2})$,
where $a(t)$ is the scale factor of the universe, are used here.
These parameters are constrained by the condition:
\begin{equation}
\label{omega_tot}
1 = \Omega + \lambda - \kappa.
\end{equation}

The lookback time from the present to the universe at
redshift $z_{1}$ is given by:
\begin{equation}
\label{lookback}
t_{0} - t_{1} = \frac{1}{H_{0}} \int_{0}^{z_{1}} (1+z)^{-1}
\lbrace (1+z)^{2} (1+\Omega_{0} z) - z(2+z)\lambda_{0} \rbrace^{-1/2} dz,
\end{equation}
and the luminosity distance $d_{L}$ by:
\begin{equation}
\label{lumdist}
d_{L} = \frac{c(1+z_{1})}{H_{0}} \times A(z)
\end{equation}
where $A(z)$ is defined as:
\[
A(z) =  \left\{
\begin{array}{ll}
          \left| \kappa \right|^{-1/2} \sinh \lbrace
                  \left| \kappa \right|^{-1/2} \int_{0}^{z_{1}}
                  \lbrace (1+z)^{2} (1+\Omega_{0} z) -
                  z(2+z) \lambda_{0} \rbrace^{-1/2} dz \rbrace
                   & (k=-1) \\
          \int_{0}^{z_{1}} \lbrace (1+z)^{2} (1+\Omega_{0} z) -
                  z(2+z)\lambda_{0} \rbrace^{-1/2} dz & (k=0) \\
          \left| \kappa \right|^{-1/2} \sin \lbrace
                  \left| \kappa \right|^{-1/2} \int_{0}^{z_{1}}
                  \lbrace (1+z)^{2} (1+\Omega_{0} z) -
                  z(2+z) \lambda_{0} \rbrace^{-1/2} dz \rbrace
                   & (k=+1) \\
\end{array}
\right.
\]
Using the luminosity distance we get the comoving volume element
$dV/d\omega dz$ subtended by solid angle $d\omega$ in the redshift range
($z$,$z+dz$):
\begin{equation}
\label{volume}
\frac{dV}{d\omega dz} = \frac{c d_{L}^{2}}{H_{0}}
                (1+z)^{-2} \lbrace (1+z)^{2}
                (1+\Omega_{0} z) - z(2+z)\lambda_{0} \rbrace^{-1/2} .
\end{equation}
\twocolumn

{}

\end{document}